\documentclass[fleqn,10pt]{wlscirep}
\usepackage[utf8]{inputenc}
\usepackage[T1]{fontenc}
\usepackage{mathptmx}
\usepackage{etoolbox}
\usepackage{color}
\usepackage{longtable}
\usepackage{siunitx}
\usepackage{comment}
\title{Ultrafast Switching in Synthetic Antiferromagnet with Bilayer Rare-Earth Transition-Metal Ferrimagnets}

\author[1,*]{Chung Ting Ma}
\author[1]{Wei Zhou}
\author[1,2]{S. Joseph Poon}
\affil[1]{University of Virginia, Department of Physics, Charlottesville, Virginia, 22904, USA}
\affil[2]{University of Virginia, Department of Materials Science and Engineering, Charlottesville, Virginia, 22904, USA}

\affil[*]{ctm7sf@virginia.edu}

\begin{abstract}
In spintronics, it is important to be able to manipulate magnetization rapidly and reliably. Several methods can control magnetization, such as by applying current pulses or magnetic fields. An applied current can reverse magnetization with nanosecond speed through the spin torque effect. For faster switching, subpicosecond switching with femtoseconds laser pulse has been achieved in amorphous rare-earth transition-metal ferrimagnets. In this study, we employed atomistic simulations to investigate ultrafast switching in a synthetic antiferromagnet with bilayer amorphous FeGd ferrimagnets. Using a two-temperature model, we demonstrated ultrafast switching in this synthetic antiferromagnet without external magnetic fields. Furthermore, we showed that if we initially stabilize a skyrmion in this heterostructure, the ultrafast laser can switch the skyrmion  state using the same mechanism. Furthermore, this bilayer design allows the control of each ferrimagnetic layer individually and opens the possibility for a magnetic tunnel junction. 
\end{abstract}
\begin{document}

\flushbottom
\maketitle

\thispagestyle{empty}

\section*{Introduction}

The ability to control magnetization is a critical component of designing memory and logical devices. In thin films, magnetizations are commonly manipulated through current or external fields. In spintronic devices, currents are often used to induce spin-transfer torque and spin-orbit torque to reliably switch magnetizations without magnetic fields \cite{Diao_2007, Manchon2019, Grimaldi2020, DuttaGupta2020, Liu2021}.  Besides electrical current, a laser pulse can also induce changes in magnetization. Subpicosecond demagnetization with femtosecond laser pulse was first observed in ferromagnetic nickel film \cite{Beaurepaire1996}. Since then, ultrafast manipulation of magnetization has drawn considerable interest for its potential applications. In ferromagnets, a multistep procedure has been demonstrated to switch magnetization \cite{Lambert2014, Medapalli2017, John2017,Hamamera2022}. For example, in FePt nanoparticles, magnetizations are first thermally demagnetized, then re-magnetized through the laser-induced inverse Faraday effect \cite{John2017}.  In antiferromagnets, optical switching of antiferromagnetic order is observed in multiferroic TbMnO$_3$ at 18 K \cite{Manz2016}. Furthermore, reliable all-optical switching of magnetization in easy-plane CrPt has been proposed by unitizing the inverse Faraday effect \cite{Dannegger2021}. Nonetheless, one-shot all-optical subpicosecond switching has only been observed in ferrimagnets, such as rare-earth transition metal (RE-TM) ferrimagnets \cite{stanciu2007,Radu2011, Ostler2012, Wienholdt2013, Graves2013, Aviles2020, Ciuciulkaite2020, vanHees2020} and recently, Mn-based crystalline alloys \cite{Davies2020}.

Amorphous RE-TM ferrimagnetic films are one of the more appealing materials for applications. They consist of two antiferromagnetically coupled RE-TM sublattices, which align in an antiparallel direction. There exists a compensation temperature (T$_{Comp}$) where the magnetic moment of the two sublattices cancel each other and magnetization goes to zero \cite{Tanaka_1987,Hansen1989}. RE-TM films contain several attractive properties, including perpendicular magnetic anisotropy (PMA) \cite{dirks1977,harris1992} and high domain wall velocity \cite{kim2017,Caretta2018}. Furthermore, they are deposited at room temperature  \cite{DING2013} and their composition can be tuned to adjust magnetization and coercivity \cite{Hansen1989,DING2013}. Recent experiments also observed skyrmions in RE-TM thin films \cite{Lee2016,woo2018,Caretta2018,Quessab2020}.  One of the most intriguing properties of RE-TM ferrimagnet is the access to one-shot all-optical ultrafast switching \cite{stanciu2007,Radu2011, Ostler2012, Wienholdt2013, Graves2013, Aviles2020, Ciuciulkaite2020, vanHees2020}. In previous studies, it is revealed that angular momentum exchange between the two different sublattices is a key ingredient in all-optical switching \cite{Ostler2012,Atxitia2012,Gridnev2018}. The requirement of having two different sublattices makes ferrimagnets, such as RE-TM, one of the few PMA materials to have this capability.  

\begin{figure}[ht]
\centering
\includegraphics[width=0.5\linewidth]{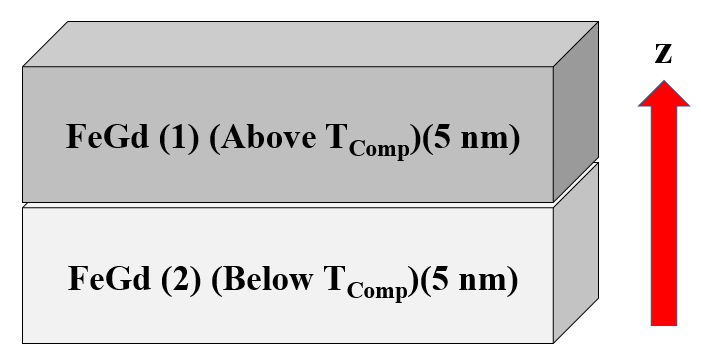}
\caption{A schematic diagram of a synthetic antiferromagnet used in this study. Two 5 nm thick FeGd combines to form the synthetic antiferromagnet. In this study, layer 1 has T$_{Comp}$ at 350 K, above room temperature, and layer 2 has T$_{Comp}$ at 250 K, below room temperature.}
\label{fig:Bilayer}
\end{figure}

In this study, we investigate laser-induced ultrafast switching in a synthetic antiferromagnet (SAF) formed from a bilayer RE-TM ferrimagnet . A schematic diagram of this heterostructure is shown in Figure \ref{fig:Bilayer}. In this heterostructure, two different compositions of 5 nm thick FeGd combine to form a 10 nm thick SAF, with one layer having T$_{Comp}$ above room temperature and the other having T$_{Comp}$ below room temperature. Such control of T$_{Comp}$ in RE/TM films has been achieved experimentally by tuning composition of Fe and Gd \cite{Hansen1989}, where a higher T$_{Comp}$ was achieved by increasing rare-earth concentration. To elaborate, this SAF arises from the cancellation of magnetization between the top and bottom FeGd layer. The magnetization in each layer is designed to be opposite but equal in magnitude at room temperature. This is obtained by choosing the T$_{Comp}$  of the top layer to be 350 K and the T$_{Comp}$  of the top layer to be 250 K. This heterostructure presents several advantages. Compared to SAF with ferromagnet or multilayer RE/TM films, SAF with RE-TM allows more flexible tuning of each layer. The thickness \cite{Hebler2016,Ma2018} and composition \cite{Hansen1989} of each layer can be varied while the net magnetization stays zero, and PMA remains robust. In contrast, SAF with ferromagnet and multilayer RE/TM films are limited in thickness and composition to maintain PMA \cite{Duine2018,Aviles2020}. Furthermore, the use of thicker layers diminishes the relative strength of interface exchange on an individual layer. This opens the possibility of switching each layer individually. In this study, we explored laser-induced ultrafast switching in SAF with RE-TM by using a two-temperature model for laser irradiation \cite{CHEN2006,Majchrzak_2015}. We found deterministic spins switching in this heterostructure, like those observed in single-layer RE-TM films. More importantly, synchronized switching are found within the same sublattice in the FeGd bilayer. Furthermore, we stabilized skyrmions in this heterostructure as initial states and found switching remains robust with a laser pulse. These findings pave the way to employ SAF with a bilayer RE-TM for spintronics applications. 

\section*{Results and Disscussions}

\begin{figure}[ht]
\centering
\includegraphics[width=\linewidth]{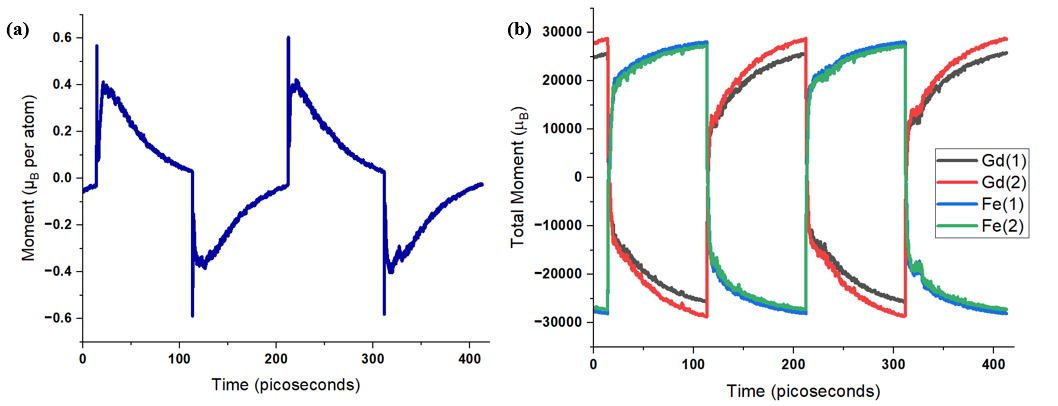}
\caption{(a) Time evolution of magnetic moment per atom with application of a 100-fs laser pulse with 30 mJ/m$^2$ fluence every 100 ps. (b) Time evolution of total magnetic moment of Gd and Fe sublattice in each FeCo layer with application of a laser pulse every 100 ps.}
\label{fig:saf1}
\end{figure}

Figure \ref{fig:saf1} shows the results of ultrafast switching in SAF with FeGd after laser pulses. Initially, the spins of Fe sublattices are pointing down (- z-direction) and the spins of Gd sublattices are pointing up (+ z-direction). The spins are initialized by the application of a 0.01 T out-of-plane magnetic field, and no external fields are applied after initialization and during the application of laser pulses. This is a stable configuration in this heterostructure as the exchange couplings between the two FeGd layers align the spins within the same sublattice parallel and the spins in different sublattices antiparallel. Also, magnetic anisotropy in FeGd holds the magnetic moment in an out-of-plane direction without an external field. After the application of a 100-fs laser pulse with 30 mJ/m$^2$ fluence, the magnetic moments in both sublattices reverse. As shown in Figure \ref{fig:saf1} (b), spins in Gd sublattices reverse from the positive to the negative direction, and spins in Fe sublattices reverse from the negative to the positive direction. Furthermore, Gd spins in both layer 1 and layer 2 reverse simultaneously and the same switching is observed in Fe spins of both layers. From Figure \ref{fig:saf1} (a), the total moment deviates from zero after the excitation by a laser pulse. This is due to the different exchange couplings and relaxation time of Fe and Gd sublattices. As discussed by previous publications \cite{Radu2011,Ostler2012,Wienholdt2013}, RE and TM sublattice have different relaxation times and lead to a transient state, where spins in RE and TM align in parallel after the first few picoseconds of a laser pulse. In this SAF, within picoseconds the initial spike in magnetic moment corresponds to the transient state, matching the previous study of single-layer RE-TM films. After the initial spike, a large downward spike is observed. This is the consequence of the high temperature from the laser pulse. As a result, the magnetic moments are not synchronized in one direction and lead to a smaller total moment. Then, as the temperature cools down, the stronger-coupled Fe sublattice is more aligned, which leads to an increase in the total moment. As the temperature cools down further, the magnetic moment begins to decrease. This is explained by the more alignment of the Gd sublattice over this period, which points opposite to the Fe atoms, and the spins begin to relax back to one of the ferrimagnetic ground states. After 100 ps, the total magnetic moment approaches back to zero. From Figure \ref{fig:saf1} (b), the moments of each sublattice are now pointing in opposite directions, corresponding to opposite spin directions from the initial configuration. Subsequence laser pulses, which were applied every 100 ps, show deterministic switching of spins in this heterostructure. 

\begin{figure}[ht]
\centering
\includegraphics[width=0.75\linewidth]{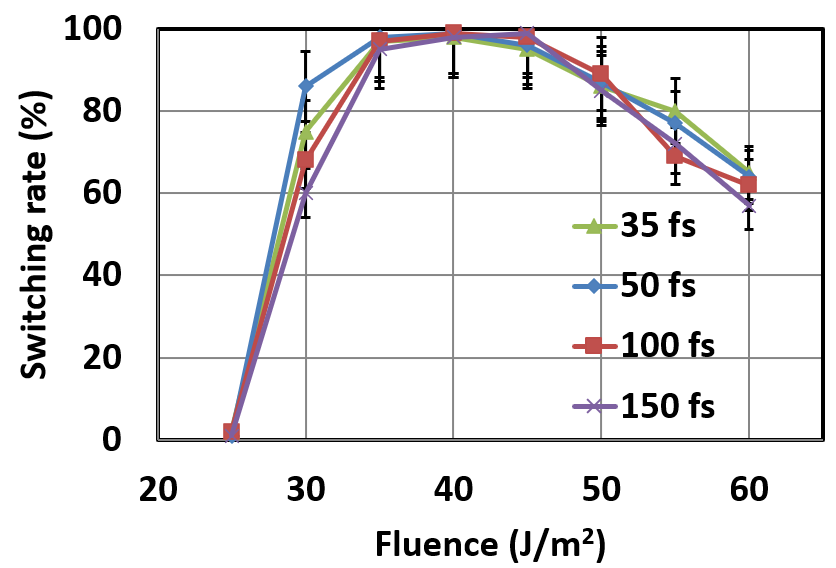}
\caption{Switching rate of magnetization in synthetic antiferromagnet with bilayer FeGd as a function of laser fluence. Simulations at each fluence is repeated 128 times and the switching rate is the percentage of switching occurred out of 128 simulations. Error bars correspond to one standard deviation from avenging. 35 fs (green), 50 fs (blue), 100 fs (red) 150 fs (purple) shows similar switching rate.}
\label{fig:rate}
\end{figure}

\begin{figure}[ht]
\centering
\includegraphics[width=\linewidth]{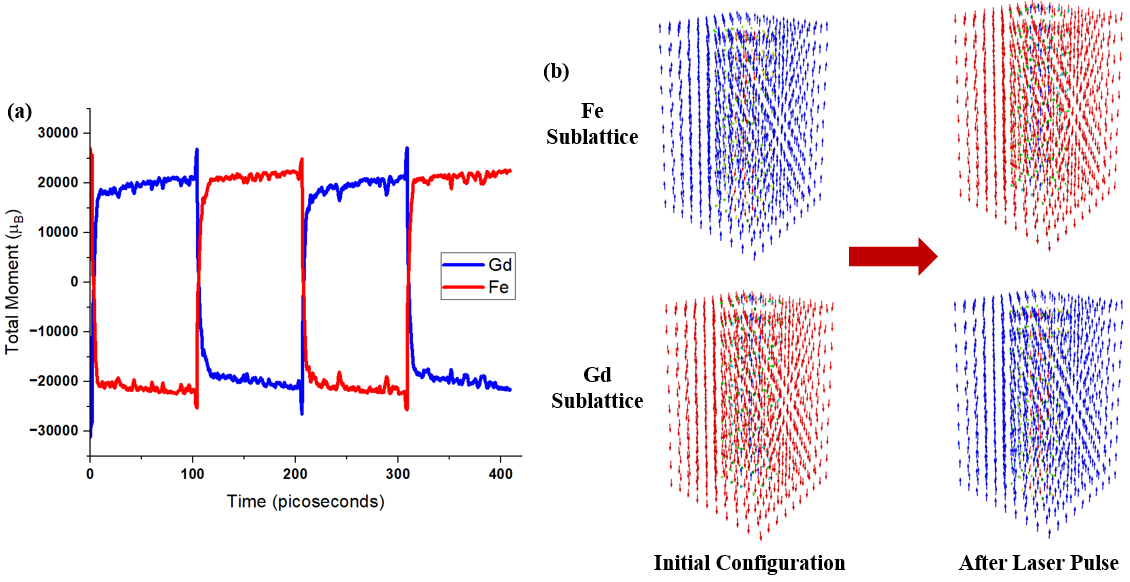}
\caption{Ultrafast switching of a skyrmion in a synthetic antiferromagnet with bilayer FeGd. (a) Time evolution of magnetic moment after application of a 100-fs laser pulse with 30 mJ/m$^2$ fluence. (b) Schematic representation of spins in Fe and Gd sublattice before and after switching by laser pulse. Red arrows represent down magnetic moments, blue arrows represents up magnetic moments, and green arrows represents near in-plane magnetic moments. Both layer of FeGd are shown here, where the top half of each schematic correponds to the top FeGd layer and the bottom half corresponds to the bottom FeGd layer.}
\label{fig:safskm}
\end{figure}

To validate the repeatability of this switching, we repeated the simulations with various laser fluence and laser pulse width. Figure \ref{fig:rate} shows the switching rate of spins in SAF with FeGd with laser fluence from 25 J/m$^2$ to 60 J/ m$^2$ and laser pulse widths of 35 fs, 50 fs, 100 fs, and 150 fs. The simulations have repeated 128 times for each set of fluence and laser pulse width, and the switching rate is the percentage of switching observations out of 128 simulations. From Figure \ref{fig:rate}, the switching rate is similar in all four laser pulse width (35 fs, 50 fs, 100 fs, and 150 fs) for various fluences. On the other hand, varying laser fluence has a significant impact on the switching rate. With a laser fluence of 25 J/ m$^2$, the switching rate is near zero. As the laser fluence increases to 30 J/ m$^2$, the switching rate increases to about 70 \%. Between laser fluence of 35 to 45 J/ m$^2$, the switching rate is above 90 \%. Above 45 J/ m$^2$, switching decreases with increases in laser fluence, reducing to about 60 \% at a laser fluence of 60 J/ m$^2$. While the mechanisms behind this dependence remain unknown, this phenomenon is likely related to the angular momentum exchange between the Fe and Gd sublattices. As discussed in other publications \cite{Ostler2012,Gridnev2018,Hebler2016}, angular momentum exchange between the two different sublattices is a crucial component of all-optical switching in RE-TM films. From intuition, at low laser fluence (< 25 J/ m$^2$), there is not enough energy to initiate the exchange of angular momentum, resulting in a zero switching rate. For high laser fluence (> 45 J/ m$^2$), an excess temperature may lead to excessive fluctuations in spins and reduces the effectiveness of angular momentum exchange between the two sublattices. While the temperature differences for different fluences certainly play a role in the switching rate, they also affect the angular momentum exchange between the two sublattices in the switching process. Further investigations are needed to reveal the underlying reasons, which are beyond the scope of this study. 

We further investigate the potential of using ultrafast switching in other magnetic states in RE-TM ferrimagnets. Figure \ref{fig:safskm} shows the ultrafast switching of a 20 nm skyrmion in SAF with FeCo. This skyrmion was initially stabilized through the interfacial Dzyaloshinskii–Moriya interaction \cite{DZYALOSHINSKY1958,Moriya1960} under 0.01 T. Details of this skyrmion calculation were discussed in previous publications \cite{Ma2019}. As seen in Figure \ref{fig:safskm} (b), initially, the spins of Fe and Gd sublattices form a skyrmion. In Fe sublattice, the spins in the core of a skyrmion are pointing in the positive direction and the spins outside are pointing in the negative direction. The spins in Gd sublattice align antiparallel to the spins in Fe sublattice. After applying a 100 fs laser pulse with 30 mJ/m$^2$ fluence, the spins in the heterostructures reverse from the initial configuration. As the spins relax, they relax back to a skyrmion configuration with spins opposite to the initial configuration, as illustrated in Figure \ref{fig:safskm} (b). From Figure \ref{fig:safskm} (a), subsequence laser pulses show the spin reversal process is repeatable.  This reversal of spin texture likely arises from the angular momentum exchange between the two different sublattices, which leads to maintaining spin texture in a subpicosecond timescale. This result demonstrates another unique feature of all-optical switching in RE-TM ferrimagnet. Furthermore, this bilayer design opens up the possibility to incorporate into a magnetic tunnel junction. One can introduce exchange bias by adding an antiferromagnetic layer on top of the top FeGd layer. By doing so, the exchange bias effect from the antiferromagnetic layer can enhance or reduce the barrier of switching in the top FeGd layer, which results in parallel or anti-parallel spins in each sublattice between the top and bottom FeGd layer. Of course, such heterostructure will need to be tested and optimized experimentally.

\section*{Conclusions}

We have performed atomistic simulations to study all-optical ultrafast switching in a 10-nm thick synthetic antiferromagnet with bilayer amorphous rare-earth transition-metal ferrimagnet. Through this study, we confirmed deterministic spin switching in the synthetic antiferromagnet by a femtosecond laser pulse. Furthermore, we demonstrated the reversal of magnetization in a skyrmion using a laser pulse.  These results indicate promise in the applications of  synthetic antiferromagnet with ferrimagnetic heterostructures in future energy-efficient high-density spintronic devices. 

\section*{Methods}

\begin{table}[ht]
\centering
\begin{tabular}{|l|l|}
\hline
Parameter & Value \\
\hline
Fe Magnetic moment ($\mu_{Fe}$)  &	2.22 $\mu_{B}$ \\
\hline
Gd Magnetic moment ($\mu_{Gd}$)	&	7.60 $\mu_{B}$ \\
\hline
Fe-Fe Exchange Interaction (J$_{Fe-Fe}$) &   2.83 x 10$^{-21}$ J \\
\hline
Gd-Gd Exchange Interaction (J$_{Gd-Gd}$)	&   1.26 x 10$^{-21}$ J \\
\hline
Fe-Gd Exchange Interaction (J$_{Fe-Gd}$)	& -1.09 x 10$^{-21}$ J \\
\hline
Anisotropy (K$_u$)	& 0.30 x 10$^{-5}$ J/m$^3$ \\
\hline
Damping ($\alpha$)	& 0.05 \\
\hline
Fe Gyromagnetic ratio ($\gamma_{Fe}$) &	1.85 T$^{-1}$s$^{-1}$ \\
\hline
Gd Gyromagnetic ratio ($\gamma_{Gd}$)    &   1.76 T$^{-1}$s$^{-1}$ \\
\hline
\end{tabular}
\caption{\label{tab:para}Parameters used for modeling magnetization dynamics in GdFe, which were obtained from Ostler el al. and Radu et al.\cite{Radu2011,Ostler2011}.}
\end{table}

\begin{figure}[ht]
\centering
\includegraphics[width=0.75\linewidth]{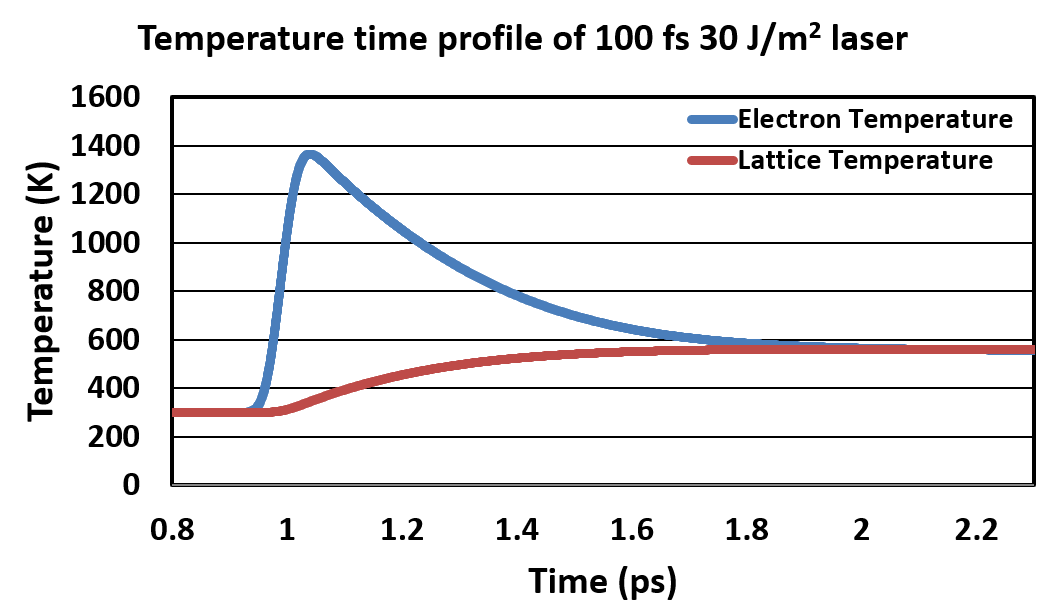}
\caption{Temporal evolution of electron and lattice temperature after irradiation of a 100-fs laser pulse with 30 mJ/m$^2$ fluence..}
\label{fig:tempearture}
\end{figure}

We built an atomistic model of Fe and Gd atoms on the FCC lattice with in-plane periodic boundary conditions. The Fe and Gd atoms are randomly distributed in the FCC lattice. For a 10 nm thick SAF, this model contains 32 x 32 x 32 sites, with half the sites (32 x 32 x 16) belonging to each layer of 5 nm thick RE-TM in the SAF. A semi-classical two-temperature model is employed to calculate the temporal evolution of electron and lattice temperature under the application of femtosecond laser irradiation \cite{CHEN2006,Majchrzak_2015}. Figure \ref{fig:tempearture} shows the temperature profile of a 100 fs laser pulse with 30 mJ/m$^2$ fluence. Within 0.1 ps, the electronic temperature reaches a peak of over 1300 K,  and the lattice temperature reaches just below 600 K. Both temperatures are above the measured Curie temperature of 540 K \cite{Ostler2011}. The atomistic spins are coupled to the electron temperature in the two-temperature model \cite{CHEN2006,Majchrzak_2015}. A stochastic Landau–Lifshitz–Gilbert (LLG) equation is used to model the magnetization dynamics \cite{GILBERT1955}. Table \ref{tab:para} summarizes the parameters used for modeling magnetization dynamics, which were obtained from Ostler et al. and Radu et al. \cite{Radu2011,Ostler2011}. The anisotropy (K$_u$) is along the z-direction. Initially, we applied an out-of-plane magnetic field to align the spins in both sublattices in the out-of-plane direction, with the Gd sublattice pointing parallel to the + z direction. To create the initial state with skyrmion, a 0.01 T magnetic field is applied \cite{Ma2019}. After initialization, the magnetic field is set to zero throughout the calculation.

\clearpage

\bibliography{ref}

\section*{Acknowledgements}

This work was supported by the DARPA Topological Excitations in Electronics (TEE) program (grant D18AP00009). The content of the information does not necessarily reflect the position or the policy of the Government, and no official endorsement should be inferred. Approved for public release; distribution is unlimited.

\section*{Author contributions statement}

C.T.M conceived the simulations and analysed the results, W.Z. and S.J.P contributed to discussions.  All authors reviewed the manuscript.

\section*{Additional information}

The author(s) declare no competing interests.

\section*{Data Availability}

The datasets generated during and/or analyzed during the current study are available from the corresponding author on reasonable request.

\end{document}